\documentclass[a4paper,12pt]{article}
\usepackage{theorem}

\newcount\driver
\newcount\bozza

\font\cs=cmcsc10 scaled\magstep1
 
scaled\magstep1                  
scaled\magstep1 
 
scaled\magstep1                 \font\indbf=cmbx10 scaled\magstep2

scaled \magstep2

{\count255=\time\divide\count255 by 60
\xdef\hourmin{\number\count255}
        \multiply\count255 by-60\advance\count255 by\time
   \xdef\hourmin{\hourmin:\ifnum\count255<10 0\fi\the\count255}}

\let\a=\alpha \let\b=\beta    \let\g=\gamma     \let\d=\delta     \let\e=\varepsilon
  \let\h=\eta      \let\k=\kappa     \let\l=\lambda
    \let\n=\nu                      \let\r=\rho
\let\s=\sigma             
\let\ps=\psi   \let\o=\omega     
 \let\D=\Delta

\def\VV{{\cal V}}
\def\WW{{\cal W}}

\def\RR{{\cal R}}\def\LL{{\cal L}}
\def\DD{{\cal D}}

\def\pp{{\bf p}}\def\xx{{\bf x}}
\def\yy{{\bf y}}\def\kk{{\bf k}}\def\nn{{\bf n}}
\def\zz{{\bf z}}

       \def\oo{{\underline \omega}}
\def\ee{{\underline \varepsilon}}

\def\u0{{\underline 0}}

\let\==\equiv

\let\io=\infty
\let\0=\noindent

\def\*{{\hfill\break\null\hfill\break}}

\def\ie{\hbox{\it i.e.\ }}

\def\tilde#1{{\widetilde #1}}

\def\la{{\langle}}
\def\ra{{\rangle}}

\def\tende#1{\,\vtop{\ialign{##\crcr\rightarrowfill\crcr
             \noalign{\kern-1pt\nointerlineskip}
             \hskip3.pt${\scriptstyle #1}$\hskip3.pt\crcr}}\,}
\def\otto{\,{\kern-1.truept\leftarrow\kern-5.truept\to\kern-1.truept}\,}

\def\wh#1{\widehat{#1}}
\def\hat#1{\wh{#1}}
\def\sqt[#1]#2{\root #1\of {#2}}

\def\hp{{\widehat \ps}}

\def\VV{{\cal V}}
\def\WW{{\cal W}}

\def\RR{{\cal R}}\def\LL{{\cal L}}
\def\DD{{\cal D}}

\def\T#1{{#1_{\kern-3pt\lower7pt\hbox{$\widetilde{}$}}\kern3pt}}
\def\VVV#1{{\underline #1}_{\kern-3pt
\lower7pt\hbox{$\widetilde{}$}}\kern3pt\,}
\def\W#1{#1_{\kern-3pt\lower7.5pt\hbox{$\widetilde{}$}}\kern2pt\,}

\def\indica{\leaders \hbox to 0.5cm{\hss.\hss}\hfill}
\def\guida{\leaders\hbox to 1em{\hss.\hss}\hfill}
\mathchardef\oo= "0521

\def\pp{{\bf p}}\def\xx{{\bf x}}
\def\yy{{\bf y}}\def\kk{{\bf k}}\def\nn{{\bf n}}
\def\zz{{\bf z}}

\def\oo{{\underline \omega}}

\def\qed{\raise1pt\hbox{\vrule height5pt width5pt depth0pt}}

\def\indic{\hbox{\raise-2pt \hbox{\indbf 1}}}

\def\ins#1#2#3{\vbox to0pt{\kern-#2 \hbox{\kern#1 #3}\vss}\nointerlineskip}

\newdimen\xshift \newdimen\xwidth \newdimen\yshift
\newcount\griglia

\def\insertplot#1#2#3#4#5#6{%
\xwidth=#1pt \xshift=\hsize \advance\xshift by-\xwidth \divide\xshift by 2%
\begin{figure}[ht]
\vspace{#2pt} \hspace{\xshift}
\begin{minipage}{#1pt}
#3 \ifnum\driver=1 \griglia=#6
\ifnum\griglia=1 \openout13=griglia.ps \write13{gsave .2
setlinewidth} \write13{0 10 #1 {dup 0 moveto #2 lineto } for}
\write13{0 10 #2 {dup 0 exch moveto #1 exch lineto } for}
\write13{stroke} \write13{.5 setlinewidth} \write13{0 50 #1 {dup 0
moveto #2 lineto } for} \write13{0 50 #2 {dup 0 exch moveto #1
exch lineto } for} \write13{stroke grestore} \closeout13
\includegraphics{griglia.ps} \fi
\includegraphics{#4.ps}\fi%
\ifnum\driver=2 \fi
\end{minipage}
\caption{#5}
\end{figure}
}

\newdimen\shift \shift=0.1truecm
\def\lb#1{%
\ifnum\bozza=1
\label{#1}\hbox{\hskip\shift$\scriptstyle#1$} \else\label{#1} \fi}

\def\be{\begin{equation}}
\def\ee{\end{equation}}
\def\bea{\begin{eqnarray}}\def\eea{\end{eqnarray}}
\def\bean{\begin{eqnarray*}}\def\eean{\end{eqnarray*}}
\def\bfr{\begin{flushright}}\def\efr{\end{flushright}}
\def\bc{\begin{center}}\def\ec{\end{center}}
\def\bal{\begin{align}}\def\eal{\end{align}}
\def\ba#1{\begin{array}{#1}} \def\ea{\end{array}}
\def\bd{\begin{description}}\def\ed{\end{description}}
\def\bv{\begin{verbatim}}\def\ev{\end{verbatim}}
\def\nn{\nonumber}
\def\Halmos{\hfill\vrule height10pt width4pt depth2pt \par\hbox to \hsize{}}
\def\pref#1{(\ref{#1})}

\begin{document}
\title{Luttinger liquid fixed point for a 2D flat Fermi surface}
\author{Vieri Mastropietro}

\maketitle
\begin{abstract}
We consider a system of 2D interacting fermions with a flat Fermi
surface. The apparent conflict between Luttinger and non Luttinger
liquid behavior found through different approximations is resolved
by showing the existence of a line of non trivial fixed points,
for the RG flow, corresponding to Luttinger liquid behavior; the
presence of marginally relevant operators can cause flow away from
the fixed point. The analysis is non-perturbative and based on the
implementation, at each RG iteration, of Ward Identities obtained
from local phase transformations depending on the Fermi surface
side, implying the partial vanishing of the Beta function.
\end{abstract}

{\cs Pacs numbers}: 71.10.Hf,71.10.Fd

\section{Introduction}

The properties of the $2D$ interacting fermions are still largely
unknown, despite the tremendous effort devoted to their understanding in the last
years. One of the most debated questions is on the possible
existence of a {\it Luttinger liquid phase}, first suggested by
Anderson [1] as an explanation of some properties of high $T_c$
superconductors, as observed also in recent experiments, see e.g.
[2] .

It has been proved, in the case of symmetric, smooth and convex
Fermi surfaces (like in the Jellium model [3] or in the Hubbard
model in the non half filled case [4]), that the wave function
renormalization $Z$ is essentially temperature independent up to
exponentially small temperatures. As in a Luttinger liquid one
expects instead a logarithmic behavior in this regime, \ie
$Z\simeq 1+O(U^2\log\b)$, such results rule out for sure the
possibility of Luttinger liquid behavior.

On the contrary the presence in the Fermi
surface of flat regions can produce non Fermi liquid behavior. The
simplest model with a flat Fermi surface is the 2D Hubbard model
at half filling, in which the Fermi surface is a square. It was
proved in [5] that the wave function renormalization is
$Z=1+O(U^2\log^2\b)$ up to exponentially small temperatures; the
presence of the $\log^2\b$ is a consequence of the Van Hove
singularities, related to the fact that the Fermi velocity is
vanishing at the corners of the squared Fermi surface, and implies
that also such a model does {\it not} show Luttinger liquid
behavior.

It is important to stress that the results in [3],[4],[5]  are
rigorous as they are based on expansions which are {\it
convergent} provided that the temperature is not too low, the
finite temperature acting as an infrared cut-offs; however such
expansions cannot give any information on the zero temperature
properties.

A lot of attention has been devoted in recent years to the zero
temperature properties of Fermi surfaces with flat regions and no
corners, which share some features with the Fermi surfaces of some
cuprates as seen in photoemission experiments. Parquet methods
results [6] and perturbative Renormalization Group (RG) analysis
[7] truncated at one loop indicate that, for repulsive
interactions, there is no indication of a Luttinger liquid phase
at zero temperature; the effective couplings flow toward a strong
coupling regime related to the onset of d-wave superconductivity.
In a more recent RG analysis truncated at 2 loops [8] one still
gets a flow to strong coupling, but in some intermediate region
some indication of Luttinger liquid behavior is found.

Apparently conflicting results are found by applying bosonization:
in [9], [10] a model of electrons on a square Fermi surface was
mapped in a collection of fermions on coupled chains, and it is
found that the correlations at zero temperature in momentum space
are similar to the one of the Luttinger model. A related but
somewhat different strategy consists in proposing an exactly
solvable 2D analogue of the Luttinger model; this approach was
pursued in [11] and [12] and again Luttinger liquid behavior up to
zero temperature was found.

A possible explanation of such conflicting results was suggested
in [13], postulating the existence for the RG flow, in addition to
the trivial fixed point associated to non interacting fermions, of
a {\it non trivial} fixed point associated to Luttinger behavior,
which could be made instable by the presence of marginally
relevant operators. In this paper we provide a quantitative
verification of such hypothesis showing explicitely the existence
of a line non trivial Luttinger fixed points for the RG flow of a
system of 2D interacting fermions with a flat Fermi surface. It is
would be not possible to derive such result directly from the
perturbative expansions, as it is related to cancellations between
graphs to all orders of the expansion which are too complex to be
seen explicitly; it is indeed well known that even in 1D Ward
Identities (WI) are necessary to prove the existence of a
Luttinger liquid fixed point [14]. Our analysis is based on the
implementation, in an exact RG approach, of WI with corrections
due the the cut-offs introduced in the multiscale analysis,
extending a technique already used to establish Luttinger liquid
behavior in a large class of 1D fermionic systems [15],[16] or 2D
spin systems [17]. Such methods are the only ones which can be
applied to {\it non exactly solvable} models, like the model
analyzed in this paper.


\section{The model}

We consider a model with a square Fermi surface similar to the one
considered in [6], [8] or [9]; the Schwinger functions are given
by functional derivatives of the generating functional
\begin{equation}\label{h}
e^{\WW(\phi)}=\int P(d\psi) e^{\VV(\psi)+\int d\xx [\psi^+_\xx\phi^-_\xx+\psi^-_\xx\phi^+_\xx]}
\end{equation}
with $\psi^\pm_\kk$ are Grassmann variables, $\kk=(k_-,k_+,k_0)$,
$k_\pm={2\pi\over L_\pm}n_\pm$, $k_0={2\pi\over\b}(n_0+{1\over
2})$, $n_\pm,n_0=0,\pm 1,\pm 2,...$ and $P(d\psi)$ is the
fermionic integration with propagator
\be\label{1a} g_\kk=\sum_{\s=\pm}\sum_{\o=\pm} {H(k_{-\s})
C_0^{-1}(\sqrt{a_0^{-2}(k_0^2+v_F^2(|k_\s|-p_F)^2)})\over-i
k_0+v_F(k_\s-\o p_F)}\equiv \sum_{\s,\o=\pm} g_{\s,\o,\kk} \ee
$H(k_{-\s})=\chi(a^{-2} k^2_{-\s})$, $\chi(t)=1$ if $t<1$ and $0$
otherwise, $C_0^{-1}(t)$ is a smooth compact support function $=1$
for $t<1$ and $=0$ for $t\ge \g$, $\g>1$. We assume, for
definiteness, $a\le {p_F\over 4},a_0\le {p_F\over 20}$ so that the
support of $g_\kk$ is over 4 disconnected regions; the Fermi
surface is defined as the set of in which $g_\kk$ for $k_0=0$ is
singular, in the limit $\b\to\io$.

By using well known properties of Grassmann integrals, see [18],
\pref{1ac1} allows to write the Grassmann field as a sum of
independent fields
\be
\psi_\kk^\pm=\sum_{\s=\pm}\sum_{\o=\pm}\psi_{\o,\s,\kk}^{\pm}\label{2a}
\ee
with $\psi^{\pm}_{\o,\s,\xx}$ independent Grassmann variables with
propagator $g_{\o,\s,\kk}$. As in [6],[8] or [9] we can consider
only interactions between parallel patches ($V=L_+ L_-$)
\be \VV=\sum_\s\sum_{\underline\o}{1\over
(V\b)^4}\sum_{\kk_1,...\kk_4}U \hat v(\kk_1-\kk_2)
\psi_{\o_1,\s,\kk_1}^{+}\psi_{\o_2,\s,\kk_2}^{-}\psi_{\o_3,\s,\kk_3}^{+}\psi_{\o_4,\s,\kk_4}^{-}
\d(\kk_1-\kk_2+\kk_3-\kk_4) \ee
with $v(\xx)$ a short range potential. The 2-point Schwinger
function is given by
\be
S_2(\xx,\yy)={\partial^2\WW(\phi)\over\partial\phi_\xx\partial\phi_\yy}|_{\phi=0}
\ee
\insertplot{300}{120}%
{.
}%
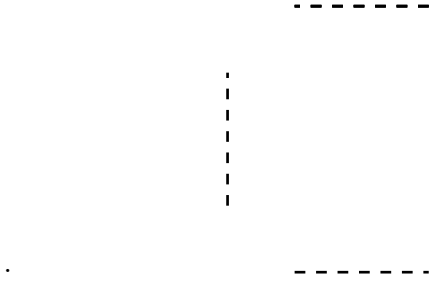
{fermi1} {\label{m3}The Fermi surface corresponding to the
singularities of $g_\kk$; the four sides are labelled by $(\s,\o)=(\pm,\pm)$.
}{0}

\section{Renormalization Group analyis}

As the interaction does not couple different $\s$ we can from now
on fix $\s=+$ for definiteness and forget the index $\s$. We
analyze the functional integral \pref{h} by performing a
multiscale analysis, using the methods of constructive Quantum Field Theory
(for a general introduction to such methods, see [18]).
The propagator \pref{1a} can be written as
sum of "single slice" propagators in the following way
\be g_\o(\xx-\yy)=\sum_{h=-\io}^0 e^{i\o p_F(x_+-y_+)}
g^{(h)}_{\o}(\xx-\yy) \label{1ac1}\ee
where
\be g^{(h)}_{\o}(\xx-\yy)= {1\over V\b}\sum_\kk
e^{i\kk(\xx-\yy)} { H(k_{-})f_h(\kk)\over-i k_0+\o v_F k_+} \ee
and $f_h(\kk)$ has support in a region $O(\g^h)$ around each flat
side of the Fermi surface, at a distance $O(\g^h)$ from it, that
is $a_0\g^{h-1}\le k_0^2+v_F^2 k^2_+\le a_0\g^{h+1}$; note that in
each term in \pref{1ac1} the change of variables $k_+\to k_++\o
p_F$ has been performed. The single scale propagator verify the
following bound, for any integer $M$
\be |g^{(h)}_{\o}(\xx)|\le C_M|{\sin a x_-\over
x_-}|{\g^h\over 1+[\g^h(|x_+|+|x_0|)]^M} \ee
The integration is done iteratively integrating out the fields
with momenta closer and closer to the Fermi surface, renormalizing
at each step the wave function. After the integration of the
fields $\psi^{(0)},...,\psi^{(h+1)}$ we obtain
\be\label{22} \int P_{Z_h}(d\psi^{(\le h)})
e^{-\VV^{(h)}(\sqrt{Z_h}\psi^{(\le h)})} \ee
where $P_{Z_h}(d\psi^{(\le h)})$ is the fermionic integration with
propagator $Z_h^{-1}(\kk) g^{(\le h)}_{\o,\kk}$, with $g^{(\le
h)}_{\o,\kk}=\sum_{k=-\io}^h g^{(k)}_{\o,\kk}$ and $Z_h$ is
defined iteratively starting from $Z_0=1$; moreover, if $\vec
p_F=(0,p_F,0)$
\bea&&\VV^{(h)}(\psi^{\le h})=\sum_{n=1}^\io
\sum_{\underline\o}{1\over (\b V)^{2n}}\sum_{\kk_1,...,\kk_{2n}}
\d(\sum_{i}\e_i (\kk_i+\o_i \vec
p_{F}))\nn\\
&& \left[\prod_{i=1}^{2n} \hat\psi^{(\le
h)\e_i}_{\o_i,\kk_i}\right] \hat
W^{(h)}_{2n}(\kk_1,...,\kk_{2n-1})\eea
By using that $\int d\kk |g^{(k)}_{\o,\s,\kk}|\le C\g^{k}$ and
$|g^{(k)}_{\o,\s,\kk}|\le C\g^{-k}$ we see that the kernels $\hat
W^{(k)}_{2n}$ are $O(\g^{-k(n-2)})$; this means that the terms
quadratic in the fields have positive scaling dimension and the
quartic terms have vanishing scaling dimension, and all the other
terms have negative dimension; we have then to properly
renormalize the terms with non-negative dimension.

Calling $\bar\kk=(k_-,0, 0)$ we define an $\LL$ operator acting
linearly on the kernels of the effective potential: \vskip.5cm
1)$\LL \hat W^{(h)}_{2n}=0$ if $n\ge 2$ \vskip.5cm 2)If $n=1$
\be\LL\hat W^h_2(\kk)=\hat W^h_2(\bar\kk)+k_0\partial_{k_0} \hat
W^h_2(\bar\kk)+ k_+
\partial_{+} \hat W^h_2(\bar\kk)\ee.
\vskip.5cm 3)
If $n=2$
\be \LL \hat W^h_4(\kk_1,\kk_2,\kk_3)= \d_{\sum_i \e_i \o_i,0}
\hat W^h_4(\bar\kk_1,\bar\kk_2, \bar\kk_3)\;. \ee
Calling $ \partial_{0} \hat W^h_2(\bar\kk)=-i z_h(k_-)$, $
\partial_{+} \hat W^h_2(\bar\kk)=\o
 z_h(k_-)$ (symmetry considerations are used)
and

$l_{h}(k_{-,1}, k_{-,2}, k_{-,3})=\hat W_4^h(\bar\kk_{1},
\bar\kk_{2}, \bar\kk_{3})$ we obtain
\bea &&\LL \VV^h={1\over\b V}\sum_{\kk}[ z_h(k_{-})\o
k_+-i k_0 z_h(k_{-})]\hat\psi^{+(\le
h)}_{\kk,\o}\hat\psi^{-(\le h)}_{\kk,\o}+\label{13}\\
&&\sum^*_{\underline\o,\underline\s}{1\over(\b
V)^4}\sum_{\kk_1,...,\kk_4} l_{h}(k_{-,1}, k_{-,2}, k_{-,3})
\hat\psi^{+(\le h)}_{\kk_1,\o_1}\hat\psi^{-(\le h)}_{\kk_2,\o_2}
\hat\psi^{+(\le h)}_{\kk_3,\o_3}\ \hat\psi^{-(\le
h)}_{\kk_4,\o_4}\d(\sum_i\e_i\kk_i))\nn \eea
where $\sum^*_{\underline\o}$ is constrained to the condition
$\sum_i \e_i \o_i \vec p_F=0$ and we have used that, by symmetry,
$W^h_2(\bar k)=0$.

We write \pref{22} as
\be \int P_{Z_h}(d\psi^{(\le h)})
e^{-\LL\VV^{(h)}(\sqrt{Z_h}\psi^{(\le
h)})-\RR\VV^{(h)}(\sqrt{Z_h}\psi^{(\le h)})} \label{33}\ee
with $\RR=1-\LL$. The non trivial action of $\RR$ on the kernel
with $n=2$ can be written as
\bea
&&\RR \hat W^h_4(\kk_1,\kk_2,\kk_3)=
[ \hat W^h_4(\kk_1,\kk_2,\kk_3)- \hat W^h_4(\bar\kk_{1},\kk_2,\kk_3)]\\
&&+[\hat W^h_4(\bar\kk_{1},\kk_2,\kk_3)-
\hat W^h_4(\bar\kk_{1},\bar\kk_{2},\kk_3)]
+ [ \hat W^h_4(\bar\kk_{1},\bar\kk_{2},\kk_3)- \hat W^h_4(\bar\kk_{1},
\bar\kk_{2},\bar\kk_{3})]\nn\eea
The first addend can be written as
\bea
&&k_{0,1}\int_0^1 dt \partial_{k_{0,1}}
\hat W_4^h(k_{-,1},k_{+,1},t k_{0,1};\kk_2,\kk_3) +\nn\\
&&k_{+,1}\int_0^1 dt \partial_{k_{+,1}} \hat W_4^h(k_{-,1}, t
k_{+,1},0;\kk_2,\kk_3) \eea
The factors $k_{0,1}$ and $k_{+,1}$ are $O(\g^{h'})$, for the
compact support properties of the propagator associated to
$\hat\psi^{+(\le h)}_{\o_1,\kk_1}$, with $h'\le h$, while the
derivatives are dimensionally $O(\g^{-h-1})$; hence the effect of
$\RR$ is to produce  a factor $\g^{h'-h-1}<1$ making its scaling
dimension negative. Similar considerations can be done for the
action of $\RR$ on the $n=1$ terms. The effect of the $\LL$
operation is to replace in $W^h_2(\kk)$ the momentum $\vec k$ with
its projection on the closest flat side of the Fermi surface.
Hence the fact that the propagator is singular over an extended
region (the Fermi surface) and not simply in a point has the
effect that the renormalization point cannot be fixed but it must
be left moving on the Fermi surface.

In order to integrate the field $\psi^{(h)}$ we can write \pref{33} as
\be
\int
P_{Z_{h-1}}(d\psi^{(\le h)}) e^{-\LL\tilde
\VV^{h}(\sqrt{Z_{h}}\psi^{(\le h)})-
\RR\VV^{(h)}(\sqrt{Z_{h}}\psi^{(\le h)})}
\ee
where $P_{Z_{h-1}}(d\psi^{(\le h)})$ is the fermionic integration
with propagator \be {1\over Z_{h-1}(\kk)} {H(k_{-}) C_h^{-1}(\kk)
\over -i k_0+\o v_F k_+}\ee
with $C_h^{-1}(\kk)=\sum_{k=-\io}^h f_k$ and
\be\label{x} Z_{h-1}(\kk)=Z_{h}(k_{-})[1+ H(k_{-})
C^{-1}_h(\kk) z_h(k _{-})] \ee
Moreover
$\LL\tilde \VV^{h}$ is the second term in \pref{13}.

We rescale the fields by rewriting the r.h.s. of \pref{33} as
\be\int P_{Z_{h-1}}(d\psi^{(\le h)}) e^{-\LL\hat
\VV^{h}(\sqrt{Z_{h-1}}\psi^{(\le h)})-
\RR\VV^{(h)}(\sqrt{Z_{h-1}}\psi^{(\le h)})}\ee
where
\be \LL\hat \VV^h(\psi)=\sum^*_{\underline\o}{1\over(\b
V)^4}\sum_{\kk_1,..,\kk_4} g_{h} (k_{-,1},k_{-,2},k_{-,3})
\hat\psi^+_{\kk_1,\o_1}\hat\psi^-_{\kk_2,\o_2}
\hat\psi^+_{\kk_3,\o_3}\hat\psi^-_{\kk_4,\o_4}\d(\sum_i\e_i
\kk_i)\label{pp} \ee
and the {\it effective couplings}
\be\label{beta2}
g_{h} (k_{-,1}, k_{-,2}, k_{-,3})=
[\prod_{i=1}^4\sqrt{Z_h(k_{-,i}) \over Z_{h-1}(k_{-,i})}]
l_{h}(k_{-,1},k_{-,2},k_{-,3})\ee
%
%
%
%
%
After the integrations of the fields
$\psi^{(0)},\psi^{(-1)},...,\psi^{(h)}$ we get an effective theory
describing fermions with wave function renormalization $Z_h$ and
effective interaction \pref{pp}. Note that $Z_h$ and $g_h$ are non
trivial functions of the momentum parallel to the Fermi surface.

We write
\be \int P_{Z_{h-1}}(d\psi^{(\le h-1)})\int
P_{Z_{h-1}}(d\psi^{(h)})e^{-\LL\hat
\VV^{(h)}(\sqrt{Z_{h-1}}\psi^{(\le h)})
-\RR\VV^{(h)}(\sqrt{Z_{h-1}}\psi^{(\le h)})} \ee
and the propagator of $P_{Z_{h-1}}(d\psi)$ is
$$\hat g^h_{\o,\s}(\kk)= H(k_{-})
{1\over Z_{h-1}(k_-)} {\tilde f_h(\kk) \over -i k_0+\o v_F k_+}$$
and
\be \tilde f_h(\kk)=Z_{h-1}(k_-) [{C^{-1}_h(\kk)\over
Z_{h-1}(\kk)}-{C_{h-1}^{-1}(\kk)\over Z_{h-1}(k_-)}] \ee
with
$H(k_{-}) \tilde f_h(\kk)$ having the same support that $H(k_{-})
f_h(\kk)$. We integrate then the field $\psi^{(h)}$ and we get
\be \int P_{Z_{h-1}}(d\psi^{(\le h-1)})e^{-
\VV^{(h-1)}(\sqrt{Z_{h-1}}\psi^{(\le h-1)})}\ee and the procedure
can be iterated. \vskip.5cm

The above procedure allows us to write $W^{(h)}_{2n}$ as a series
in the effective couplings $g_k$, $k\ge h$, which is {\it
convergent}, see [18], provided that $L_-$ is finite and
$\e_h=\sup_{k\ge h}||g_k||$ small enough; moreover
$||W_{2n}||=O(\g^{-h(n-2)})$. A similar analysis can be repeated
for the 2-point function.

However even if the couplings $g_k$ starts with small values, they
can possibly increase iterating the RG and at the end reach the
boundary of the (estimated) convergence domain; if this happen,
all the above procedure looses its consistency. A finite
temperature acts as an infrared cut-off saying that the RG has to
be iterated up to a maximum scale $h_\b=O(\log\b)$ and, up to
exponentially small temperatures \ie $\b\le O(e^{\k |U|^{-1}})$,
then surely the effective couplings are in the convergence domain;
however, in order to get lower temperatures, more information on
the effective couplings are necessary.

\section{RG flow and Luttinger liquid fixed point}

The RG analysis seen in the previous section implies that the
effective coupling $g_h$ verify a flow equation of the form
\be\label{d}
g_{h-1}=g_h+\b_{g}^{(h)}(g_h;...;g_0)
\ee
where the r.h.s. of the above equation is called {\it Beta
function}, which is expressed by a convergent expansion in the
couplings if $\e_h$ is small enough. The first non trivial
contribution to $\b_g^{(h)}$, called $\b_g^{(2)(h)}$, is quadratic
in the couplings and it is given by
\be \b_{g}^{(2)(h)}=\b_h^{(a)}+\b^{(b)}_h\ee
where
\bea
&& \b_h^{(a)}=\int d\pp H(k_{1,-}-p_-)H(k_{3,-}+p_-)g_h(k_{1,-},k_{1,-}-p_-,k_{3,-})\label{nbf}\\
&&g_h(k_{1,-}-p_-,k_{2,-},k_{3,-}+p_-)
{f_h(\pp)  C_h(\pp)\over p_0^2+v_F^2 p^2_+}\nn\\
&&\b_h^{(b)}=-\int d\pp
H(k_{2,-}-p_-)H(k_{3,-}-p_-)g_h(k_{1,-},k_{2,-}-p_-,k_{3,-}-p_-)\nn\\
&&g_h(k_{2,-}-p_-,k_{2,-},k_{3,-}) {f_h(\pp)  C_h(\pp)\over
p_0^2+v_F^2 p^2_+}\label{22oo}\eea
The above expression essentially coincides with the one found in
[6] or [8]; it is indeed well known that the lowest order
contributions to the Beta function are essentially independent by
RG procedure one follows.

The flow equation \pref{d} encodes most of the physical properties
of the model, but its analysis is extremely complex. Some insights
can be obtained by truncating the beta function at second order,
and by the numerical analysis of the resulting flow by
discretization of the Fermi surface; it is found, see [6] or
[8],that $g_h(k_{-,1},k_{-,2},k_{-,3})$ has a flow which, for
certain values of $k_{-,1},k_{-,2},k_{-,3}$ increases and reach
the estimated domain of convergence of the series for $W_n^{(k)}$.
While this increasing can be interpreted as a sign of instability,
mathematically speaking this means that the truncation procedure
becomes inconsistent.

A basic question is about the fixed points of the flow equation
\pref{d}; in particular if there is,in addition to the trivial
fixed point $g_h=0$, a non trivial fixed point corresponding to
Luttinger liquid behavior. Note first that the set \be\label{is}
g_{h}(k_{-,1},k_{-,2},k_{-,3})={1\over
L_-}\d(k_{1,-}-k_{2,-})\l_{h} \ee
with
$\l_h$ constant in $\kk$, is invariant under the RG flow, in the sense
that if $\LL\VV^{(k)}$ has the form,for $k\ge h$
\be\label{ggg}
{1\over(\b
V)^3}\sum_{\kk,\kk',\pp\atop\o,\o'} \l_h\d_{p_-,0}
\hat\psi^{+}_{\o,\kk}
\hat\psi^-_{\o,\kk+\pp}\hat\psi^{+}_{\o',\kk'}
\hat\psi^-_{\o',\kk'-\pp} \ee
the same is true for $\LL\VV^{(h-1)}$. This can be checked by the
graph expansion. In the graphs contributing to $W_4^{(h)}$, the
external lines of the graphs contributing to $W_4^{(h)}$ either
comes out from a single point, or are connected by a chain of
propagators with the same $\o,k_-$. Moreover in each Feynman graph
the only dependence from the momenta of the external lines is
through the function $H(k_-)$ which are $1$ in the support of the
external fields $\int d\kk H(k_-)\psi^\pm_\kk=\int d\kk
\psi^\pm_\kk$. For the same reasons also $Z_h$ is independent from
$k_-$.

The crucial point is that, in the invariant set \pref{is}, some
dramatic cancellation are present implying the following {\it
asymptotic vanishing of the beta function} (which will be proved
in the subsequent sections)
\be \b^{(h)}_g=O(\g^h\e_h^2)\label{cd11} \ee
saying that there is a {\it cancellations} between
the graphs with four external lines and the graphs with two lines
contributing to the square of $Z_h$ , see \pref{beta2}; such
graphs are $O(1)$ but there are cancellations making the size of
the sum of them $O(\g^h)$. At the second order \pref{cd11} can be verified
from \pref{22oo} and \pref{is}; at third order, it is compatible
with (A16),(A15) and (4.8) of [8].

The validity of \pref{cd11} immediately implies the existence of a
line of {\it non-trivial fixed points} for \pref{d} of the form
\be\label{fp} g_{-\io}(k_{-,1},k_{-,2},k_{-,3})={1\over
L_-}\d(k_{1,-}-k_{2,-})\l_{-\io} \ee
with $\l_{-\io}$ $\kk$-independent and continuous function of $U$,
$\l_{-\io}=\l_0+O(U^2)$, $\l_0=c U$ for a suitable constant $c$.

Note also that to such fixed point is associated Luttinger liquid
behavior, as, from \pref{x}, $Z_h\simeq \g^{2\h h}$, with $\h=a
\l_0^2+O(U^3)$ and $\bar\pp= (0,\bar p_+,\bar p_0)$
\bea &&a=\lim_{h\to-\io}\sum_{\o'}{\l_0^2\over L_-^2} {1\over
h}\int d\kk' d\bar p_+ d\bar p_0 {H(k'_-)C_h(\kk')\over-i k'_0+\o'
v_F
k'_+}\\
&&{H(k'_-)C_h(\kk'+\bar\pp)\over-i (k'_0+\bar p_0)+\o'
v_F(k'_++\bar p_+)} {\partial\over
\partial k_+} {H(k_-) C_h^{-1}(\kk-\bar\pp)\over -i (k_0-\bar p_0)+\o v_F (k_+-\bar
p_+)}|_{k_0=k_+=0}\nn \eea

Indeed the 2-point Schwinger function can be written as
\bea S_2(\xx,\yy)=\sum_{\o=\pm} e^{i\o p_F(x_+-y_+)}
{1\over V\b}\sum_\kk e^{i\kk(\xx-\yy)}
{g^{(h)}_{\o}(\kk)\over Z_h}[1+A^{(h)}(\kk)]\nn \eea
with $A^{(h)}(\kk)=O(\e_h)$, so that
\bea && S_2(\xx,\yy)=\sum_{\o=\pm}e^{i\o p_F(x_+-y_+)}\\
&&{1\over V\b}\sum_\kk e^{i\kk(\xx-\yy)} {H(k_{-})
C_0^{-1}(\kk)\over -i k_0+\o v_F k_+}{1+A(\kk)\over |k_0^2+v_F^2
k_+^2|^\h}\nn \eea
with $|A(\kk)|\le C |U|$. This means that to the fixed point is
associated Luttinger liquid behavior, as the wave function
renormalization vanishes at the Fermi surface as a power like with
a non-universal critical index;the Luttinger liquid behavior is
found only if $L_-$ is finite, as if $L_-\to\io$ the critical
index is vanishing. Note also that the cancellation in \pref{cd11}
reduce to the one in 1D if $k_-=k'_-$ in \pref{ggg}.

\section{The auxiliary model}

There is essentially no hope of proving a property like
\pref{cd11} directly from the graph expansion, as the algebra of
the graphs is too cumbersome (except than at one loop in which it
is easy to check). We will follow instead the same strategy for
proving the asymptotic vanishing of the Beta function in 1D
followed in [15],[16], considering an {\it auxiliary model} with
the same beta function, up to irrelevant terms, but verifying
extra symmetries, from which a set of Ward Identities can be
derived. In the present case, such identities are related to the
invariance under local phase transformations depending on the
Fermi surface side, which in the model \pref{h} is broken by the
cut-off function $C_0^{-1}(\kk)$ and by the lattice.

We consider an {\it auxiliary model} with generating function is
\be\label{sax} \int \DD\psi \prod_{\o}e^{\int d\kk
H^{-1}(k_-)C_{h,N}(\kk)(-i k_0+\o k_+ v_F)
\hat\psi^+_{\o,\kk}\hat\psi^-_{\o,\kk}+
\bar\VV(\psi)+\sum_{\e=\pm}\int d\xx
\psi^{\e}_{\o,\xx}\phi^{-\e}_{\o,\xx}+\int d\xx J_{\xx}
\r_{\o,\xx}}
 \ee
with $C^{-1}_{h,N}(\kk)=\sum_{k=h}^N  f_k(k_0^2+v_F^2 k_+^2)$,
$h\le 0$ $\r_{\o,\xx}=\psi^+_{\xx,\o}\psi_{\xx,\o}$
\bea\label{qa} &&\bar\VV={U \over L_-}\sum_{\o,\o'}\int d\xx dy_-
v(x_0-y_0,x_+-y_+) \psi^+_{\o,x_-,x_+,x_0}\psi^-_{\o,x_-,x_+,x_0}
\psi^+_{\o',y_-,x_+,x_0}
\psi^-_{\o',y_-,x_+,x_0}\nn\\
&&={U\over L_-}{1\over(\b V)^2}\sum_{\kk,\kk'}{1\over \b
L_+}\sum_{p_0,p_+} \hat v(p_0,p_+) \hat\psi^{+}_{\o,\kk}
\hat\psi^-_{\o,\kk+(0,p_+,p_0)}\hat\psi^{+}_{\o',\kk'}
\hat\psi^-_{\o',\kk'-(0,p_+,p_0)}\eea
with $v(x_0,x_+)$ a short range interaction. The above functional
integral is very similar to the previous one, with the difference
that there is an ultraviolet cut--off $\g^N$ on the $+$ variables ,
which will be removed at the end, and an ultraviolet cut-off
$O(1)$ on the $-$ variables; such features are present also in the
models introduced in [9] or [12].

Again \pref{sax} can be analyzed by a multiscale integration based
on a decomposition similar to \pref{1ac1}, with the difference
that the scale are from $h$ to $N$.

In the integration of the scales between $N$ and $0$, the {\it
ultraviolet scales}, there is no need of renormalization;
apparently the terms with two or four external lines have positive
or vanishing dimension but one can use the non locality of the
interaction to improve their scaling dimension. We integrate (with
$\LL=0$) the fields $\psi^{(N)},\psi^{(N-1)},..,\psi^{(k)}$ and we
call $W^{(k)}_{2n,m}$ the kernels in the effective potential
multiplying $2n$ fermionic fields and $m$ $J$ fields. Again the
dimension is $\g^{-k(n+m-2)}$, $k\ge 0$ and we have to
improve the bounds using the non-locality of the interaction.  We
can write
\bea && W^{(k)}_{2,0}(\xx,\yy)=\int d\yy_1  U
{v(x_0-y_{1,0},x_+-y_{1,+}) \over L_-}
W^{(k)}_{0,1}(\yy_1)g^{(k,N)}(\xx-\yy_2)
W^{(k)}_{2,0}(\yy_2;\yy)\nn\\
&&+U\int d\yy_2 {v(x_0-y_{0,1},x_+-y_{+,1})\over L_-}
g^{(k,N)}(\xx-\yy_2)
W^{(k)}_{2,1}(\yy,\yy_2;\yy_1)+\label{db1o00}\\
&&U\d(\xx-\yy)\int d\yy_1 {v(x_0-y_{1,0},x_+-y_{1,+})\over L_-}
W^{(k)}_{0,1}(\yy_1) \nn \eea
\insertplot{300}{85} {\ins{130pt}{60pt}{$+$}
\ins{50pt}{60pt}{$=$}
\ins{280pt}{60pt}{$+$}
} {verticiT11}
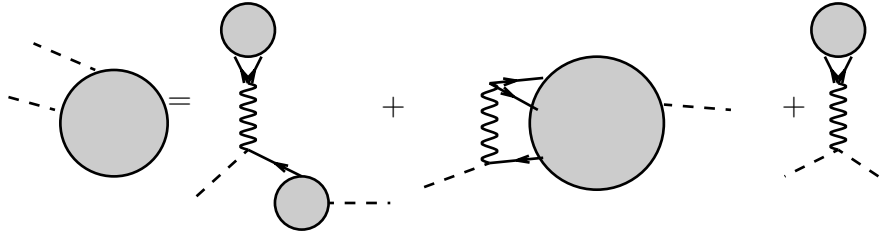
{\label{h2} Graphical representation of \pref{db1o00}; the blobs
represent $W^{(k)}_{n,m}$, the wiggly lines represent $v$, the
lines $g^{(k,N)}$} {0}

The first and the third addend of Fig.2 are vanishing, by the
symmetry $g(k_0,k_+,k_-)=-g(-k_0,-k_+,k_-)$; hence, using that
$\|g^{(j)}\|_1\le \tilde C\g^{-j}$ and that $W_{2,1}^{(k)}$ is $O(U)$
(by induction), we obtain the following bound
\be \|W^{(k)}_{2,0}\|\le C{|U|\over L_-}
\|W^{(k)}_{2,1}\|\cdot \sum_{j=k}^N \|g^{(j)}\|_1 \le
C|U| \g^k L_-^{-1}\g^{-2 k} \label{x3} \ee
Note that
we have a gain $O(L_-^{-1}\g^{- 2k)})$, due to the fact that we are integrating over a fermionic
instead than over a bosonic line.

Similar arguments can be repeated for $W^{(k)}_{0,2}$, which can
be decomposed as in Fig 3. \insertplot{300}{100} {\ins{150pt}{60pt}{$+$}
\ins{270pt}{60pt}{$+$}
} {verticiT15x} {\label{h4} Decomposition of
$W^{(k)}_{2,0}$ }{0} The second term in the figure is bounded by
$O(|U| L_-^{-1} \g^{-2k})$. A similar bound is found for the third
term in Fig.3; regarding the first term, we can rewrite it as
\bea &&\int d\xx d\bar\zz
[g^{(k,N)}(\zz-\xx)]^2 {U\over L_-} v(x_0-\bar z_0, x_+-\bar z_+) W^{(k)}_{0,2}(\bar\zz,\yy)=\nn\\
&&\int d\xx d\bar\zz
U {v({\bar z}_0-z_0,\bar z_+ -z_+)\over L_-}[g^{(k,N)}(\xx-\zz)]^2 W^{(k)}_{0,2}(\bar\zz,\yy)\\
&&+\int d\xx d\bar\zz { U\over L_-}[v(\bar z_0-x_0,\bar
z_+-x_+)-v(\bar z_0-z_0,\bar
z_+-z_+)][g^{(k,N)}(\xx-\zz)]^2W^{(k)}_{0,2}(\bar\zz,y_0)\nn
\label{l11} \eea
and using that
\be\int d\xx [g^{(k,N)}(\xx-\zz)]^2=\int dk_- H(k_-)
\int
dk_{0}dk_{+}{C^{-2}_{k,N}(\kk)\over (-i k_0+k_+)^2}=0 \label{jg}
\ee
the first addend is vanishing; the second addend, by using the
interpolation formula for $v(\bar z_0-x_0,\bar z_+-x_+)-v(\bar
z_0-z_0,\bar z_+-z_+)$,
%
%
can be bounded by $C |U| \g^{-k}$, as by induction
$||W^{(k)}_{0,2}||\le C|U|$. A similar analysis proves the bound
for $W^{k)}_{4,0}$.

After the integration of the fields
$\psi^{(N)},\psi^{(N-1)},...,\psi^{(-1)}$ we get a Grassman
integral very similar to \pref{h}; the integration of the
remaining fields $\psi^{(0)},\psi^{(-1)},..$ is done following the
same procedure as in section 3, with the effective coupling of the
form \pref{is} and $\LL\VV^{(k)}$ of the form \pref{ggg}. The
crucial point is that the beta function {\it coincides} with the
beta function for the model \pref{h} up to $O(\g^h)$ terms; hence
it is enough to prove the validity of \pref{cd11} in the auxiliary
model.

\section{Ward Identities}

We derive now a set of Ward Identities relating the Schwinger
functions of the auxiliary model \pref{sax}; by performing the change of
variables
\be \psi^\pm_{\o,\xx}\to e^{\pm i\a_{\o,\xx}}\psi^{\pm}_{\o,\xx}
\ee
and making a derivative with respect to $\a_{\xx,\o}$ and to the
external fields we obtain
\bea\label{sax1} &&\int d\kk' [H^{-1}(k'_-+p_-)C_{h,N}(\kk'+\pp)
(-i (k'_0+p_0)+\o v_F (k'_++p_+))-\nn\\
&&H^{-1}(k'_-)C_{h,N}(\kk') (-i k'_0+\o v_F k'_+)]\la
\hat\psi^+_{\o,\kk'+\pp}
\hat\psi^-_{\o,\kk'}\hat\psi^+_{\o',\kk-\pp} \hat\psi^-_{\o',\kk}\ra=\nn\\
&&\d_{\o,\o'}[\la \hat\psi^+_{\o',\kk-\pp}
\hat\psi^-_{\o',\kk-\pp}\ra-\la \hat\psi^+_{\o',\kk}
\hat\psi^-_{\o',\kk}\ra] \eea
where $\la \hat\psi^+_{\o,\kk'+\pp}
\hat\psi^-_{\o,\kk'}\hat\psi^+_{\o',\kk-\pp}
\hat\psi^-_{\o',\kk}\ra$ is the derivative with respect to
$J_\pp,\phi^+_{\o',\kk-\pp},\phi^-_{\o',\kk}$ of \pref{sax}.
Computing \pref{sax1} for $p_-=0$ we get,if $\bar\pp=(0,\bar
p_+,\bar p_0)$
\bea\label{47} &&(-i \bar p_0+\o v_F \bar p_+)
\la\r_{\bar\pp,\o}\hat\psi^+_{\kk,\o'}
\hat\psi^-_{\kk-\bar\pp,\o'}\ra+\D(\kk,\bar\pp)=\nn\\
&&\d_{\o,\o'}[\la \hat\psi^+_{\o',\kk-\bar\pp}
\hat\psi^-_{\o',\kk-\bar\pp}\ra -\la \hat\psi^+_{\o',\kk}
\hat\psi^-_{\o',\kk}\ra] \eea
and
\be \D(\kk,\bar\pp)=\int d\kk'
C(\kk',\bar\pp)\la\hat\psi^+_{\o,\kk'+\bar\pp}
\hat\psi^-_{\o,\kk'}\hat\psi^+_{\o',\kk-\bar\pp}\hat\psi^-_{\o',\kk}\ra\ee
with
\bea &&C(\kk,\bar\pp)=(-i k_0+\o v_F k_+)\\
&&[C_{h,N}(\kk+\bar\pp)-C_{h,N}(\kk)]+(-i \bar p_0+\o v_F \bar
p_+) [C_{h,N}(\kk+\bar\pp)-1]\nn \eea
In deriving the above equation we have used that
\bea &&\int d\kk' H^{-1}(k'_-)\la\hat\psi^+_{\o,\kk'+\bar\pp}
\hat\psi^-_{\o,\kk'}\hat\psi^+_{\o',\kk-\bar\pp}\hat\psi^-_{\o,\kk}\ra=\nn\\
&&\int d\kk' \la\hat\psi^+_{\o,\kk'+\bar\pp}
\hat\psi^-_{\o,\kk'}\hat\psi^+_{\o',\kk-\bar\pp}\hat\psi^-_{\o',\kk}\ra\eea
for the compact support properties of the fields $\psi_{\kk}$ and
$H^2=H$; note also the crucial role of the condition $p_-=0$ in
the above derivation.


The presence of the term $\D(\kk,\bar\pp)$ in the Ward Identity
\pref{47} is related to the presence of the ultraviolet cut-off;
as in 1D, such a term is not vanishing even in the limit $N\to\io$
and it is responsible of the anomalies, see [15],[16]. The
following correction identity holds, similar to the one in the 1D
case
\be\label{cd1} \D(\kk,\bar\pp)= \n (-i\bar p_0-\o v_F
\bar p_+)\sum_{\o''=\pm}\la\r_{\bar\pp,\o''}\hat\psi^+_{\o',\kk-\bar\pp}
\hat\psi^-_{\o',\kk}\ra+R^{2,1}_{\o}(\kk,\bar\pp) \ee
with $R^{2,1}_{\o}$ a small correction. Indeed $R^{2,1}_{\o}$ can
be written as functional derivative, with respect to
$\phi^+,\phi^-,J$, of
\be e^{\WW_{\D} (J,\phi)}= \int P(d\psi)e^{-V(\psi)+\sum_\o\int
d\zz
[\psi^+_{\o,\zz}\phi^-_{\o,\zz}+\phi^+_{\o,\zz}\psi^-_{\o,\zz}]+
T_0(J,\psi)-T_{-}(J,\psi)}\label{fgbnn}\ee
with
\be T_0(\psi)= \int {d\bar p_+\over (2\pi)}{d\bar p_0\over (2\pi)}
{d\kk\over (2\pi)^3} C(\kk,\bar\pp) J_{\bar\pp}
\hp^+_{\kk+\bar\pp,\o}\hp^-_{\kk,\o}\ee \be
T_-(\hat\psi)=\sum_{\o'}\int {d\bar p_+\over (2\pi)}{d\bar
p_0\over (2\pi)}
 {d\kk\over (2\pi)^3} \n J_{\bar\pp}
(-i \bar p_0-\o v_F \bar p_+)
\hp^+_{\kk+\bar\pp,\o'}\hp^-_{\kk,\o'}\label{lalla78} \ee
\pref{fgbnn} can be evaluated by a multiscale integration similar
to the previous one, the only difference being that $\int J \r$ is
replaced by $T_0-T_-$. The terms with vanishing scaling dimension
of the form $J\psi^+\psi^-$ can be obtained from the contraction
of $T_0$ and $T_-$; in the first case we can perform a
decomposition similar to the one in Fig.3, see Fig. 4.
\insertplot{300}{90} {\ins{150pt}{60pt}{$+$}
\ins{270pt}{60pt}{$+$}
} {verticiT15xx}
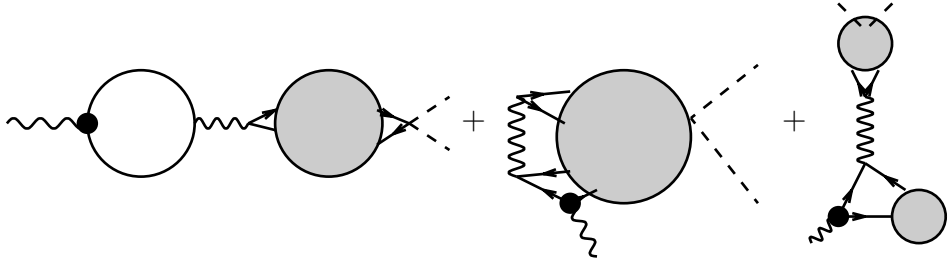
{\label{m9} Terms obtained from the contraction of $T_0$; the
black dot represents $C(\kk,\bar\pp)$} {0}
Regarding the second and third term, we can proceed exactly as in
the previous section, the main difference being that at least one
of the two fields have scale $N$ so that they obey to the bound
$O(\g^{-2k}\g^{-(1/2)(N-k)})$. This follows from the fact that
when $C$ is multiplied by two propagators we get
\bea &&C(\kk,\bar\pp)g^{(i)}(\kk)
g^{(j)}(\kk+\bar\pp)={f_i(\kk)\over -i k_0+\o v_F k_+}
[{f_j(\kk+\bar\pp) \over
C^{-1}_{h,N}(\kk+\bar\pp)}-f_j(\kk+\bar\pp)]\nn\\
&&-{f_j(\kk+\bar\pp)\over -i (k_0+\bar p_0)+\o v_F (k_++\bar p_+)}
[{f_i(\kk) \over C^{-1}_{h,N}(\kk)}-f_i(\kk)] \eea
which is non vanishing only if one among $i$ or $j$ are equal to $h$ or $N$.

The main difference with the analysis in the previous section is
that in the first term of Fig 4; the "bubble" in Fig. 3 was
vanishing, while here it is not. We choose $\n$ in \pref{fgbnn}
equal to the value of this bubble, in order to cancel it. The
value of the bubble is given by
\be \n=U v(\bar p_0,\bar p_+)\int {d\kk\over
(2\pi)^3}{C(\kk,\bar\pp)\over -i\bar p_0 -\o v_F\bar p_+}
g_\o^{(\le N)}(\kk) g_\o^{(\le N)}(\kk+\bar\pp) \ee
and,in the limit ${N\to\io}$, $\n=U v(\bar p_0,\bar p_+){a\over
4\pi^2}$.
%
%
%
%
%
%
Hence for $k_0,k_+=O(\g^h)$
\be|R^{2,1}_\o(\kk,\bar\pp)|\le C\e_h  ^2 \g^{-2 h}\ee
\section{Schwinger-Dyson equation}
An immediate consequence of the analysis in the previous section
is that, for momenta computed at the infrared scale
$|\kk|=|\kk'|=|\kk+\hat\pp|=|\kk'-\hat\pp|=|\hat\pp|=\g^h$
\bea\label{44} &&\la \hat\psi^{-}_{\o,\kk}
\hat\psi^+_{\o,\kk+\hat\pp} \hat\psi^{-}_{\o',\kk'}
\hat\psi^+_{\o',\kk'-\hat\pp}
\ra =\nn\\
&&{1\over (Z_h)^2} {\l_h\over L_-} g^{(h)}_{\o,\kk}
g^{(h)}_{\o',\kk'}g^{(h)}_{\o,\kk'+\hat\pp}g^{(h)}_{\o',\kk'-\hat\pp}
(1+O(\e_h))\nn\\
&&\la \hat\psi^{-}_{\o,\kk}
\hat\psi^+_{\o,\kk}\ra={g^{(h)}_{\o,\kk}\over Z_h}(1+O(\e_h)) \eea
This says that relations between the effective couplings at a
certain scale $h\le 0$ can be obtained from relations between the
Schwinger functions of the auxiliary model \pref{sax} computed at
the infrared cut-off scale. The starting point for deriving such
relations is the Schwinger-Dyson equation for the 4-point
function, given by, if the external momenta are computed at the
infrared scale and $\hat\pp=(0,\hat p_+,\hat p_0)$  and
$\bar\pp=(0,\bar p_+,\bar p_0)$
\bea\label{sd} &&\la\hat\psi^{-}_{\o,\kk}
\hat\psi^+_{\o,\kk-\hat\pp} \hat\psi^{-}_{\o',\kk'}
\hat\psi^+_{\o',\kk'+\hat\pp} \ra=\label{sd}\nn\\
&& \sum_{\o''}\{{U\over L_-} v(\hat p_0,\hat p_+)
g_{\o',\kk'+\hat\pp} \la \hat\psi^{-}_{\o',\kk'} \hat\psi^+_{\o',\kk'}
\ra\la \r_{\hat\pp,\o''}\hat\psi^{-}_{\o,\kk}
\hat\psi^+_{\o,\kk-\hat\pp} \ra
+\\
&&{U\over L_-}g_{\o',\kk'+\hat\pp} \int {d\bar p_0\over
(2\pi)}{d\bar p_+\over (2\pi)}
 v(\bar p_0,\bar p_+)\la \r_{\bar\pp,\o''}
\hat\psi^{-}_{\o,\kk}
\hat\psi^+_{\o,\kk-\hat\pp}\hat\psi^{-}_{\o',\kk'}
\hat\psi^+_{\o',\kk'+\hat\pp-\bar\pp}
 \ra\}\nn
 \eea
\insertplot{250}{130} {





%

\ins{95pt}{75pt}{$=$}

%







%

\ins{200pt}{75pt}{$+$}

%






\ins{235pt}{70pt}{$\pp$}
} {dyson} {\label{m9}
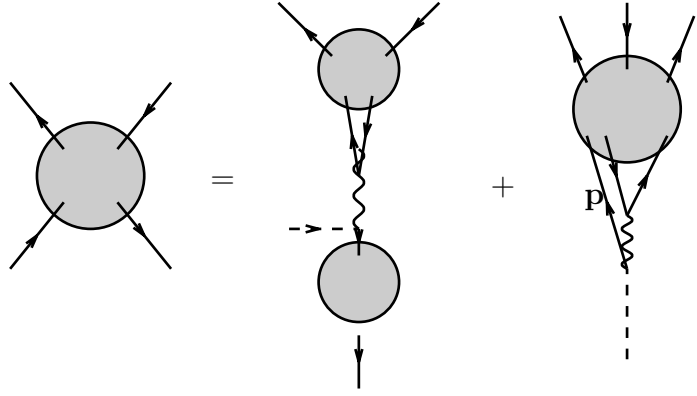
Graphical representation of \pref{sd}; the dotted line represent
the free propagator} {0}

By the WI \pref{47},\pref{cd1}
\bea\label{45} &&(-i\hat p_0+\o' v_F \hat p_+)\bar
\la\r_{\hat\pp,\o'}\hat\psi^-_{\kk,\o} \hat\psi^+_{\kk-\hat\pp,\o}\ra=\nn\\
&&A_{\o,\o'}(\pp)[\la \hat\psi^-_{\o,\kk-\hat\pp}
\hat\psi^+_{\o,\kk-\hat\pp}\ra -\la \hat\psi^-_{\o,\kk}
\hat\psi^+_{\o,\kk}\ra]+H^{(2,1)}_{\o,\o'}(\kk,\pp)\eea
with $A_{\o,\o}(\pp)=1+O(\e_h)$, $A_{\o,-\o}(\pp)=O(\e_h)$ and
even in $\pp$; moreover $H_{\o,\o'}(\kk,\pp)$ is a linear
combination of the $R^{2,1}$ functions in \pref{cd1} with bounded
coefficients.
%
%
The WI for the 4 point function is given by
\bea &&(-i\bar p_0+\o v_F \bar p_+)\la\r_{\bar\pp,\o}
\hat\psi^{-}_{\o,\kk}
\hat\psi^+_{\o,\kk-\hat\pp}\hat\psi^{-}_{-\o,\kk'}
\hat\psi^+_{-\o,\kk'+\hat\pp-\bar\pp}
\ra=\label{cd1212}\\
&& \la \hat\psi^-_{\kk-\bar\pp,\o}\hat\psi^+_{\kk-\hat\pp,\o}
\hat\psi^-_{\kk', -\o} \hat\psi^+_{\kk'+\hat\pp-\bar\pp,-\o} \ra
-\la \hat\psi^-_{\kk,\o}\hat\psi^+_{\kk-\hat\pp+\bar\pp,\o}
\hat\psi^-_{\kk', -\o} \hat\psi^+_{\kk'+\hat\pp-\bar\pp,-\o}
\ra\nn\\
&&+\n\sum_{\o'}(-i\bar p_0-\o v_F \bar p_+)\la
\r_{\bar\pp,\o}\hat\psi^-_{\kk,\o} \hat\psi^+_
{\kk-\hat\pp,\o}\hat\psi^-_{\kk', -\o}
\hat\psi^+_{\kk'+\hat\pp-\bar\pp,-\o}\ra
+R_{\o}^{4,1}(\kk,\kk',p_0)\nn \eea
and similar ones, so that the second addend of the l.h.s. is given
by
\bea &&\int d\bar p_0d\bar p_+ \chi_\e(\bar\pp){\hat v(\bar
p_0,\bar p_+)\over -i \bar p_0+\o'' v_F\bar p_+}\big [A_1(\bar\pp)
\la \hat\psi^-_{\kk-\bar\pp,\o}\hat\psi^+_{\kk-\hat\pp,\o}
\hat\psi^-_{\kk', \o'} \hat\psi^+_{\kk'+\hat\pp-\bar\pp,\o'}
\ra\nn\\
&&+A_2(\bar\pp)\la
\hat\psi^-_{\kk,\o}\hat\psi^+_{\kk-\hat\pp+\bar\pp,\o}
\hat\psi^-_{\kk', \o'} \hat\psi^+_{\kk'+\hat\pp-\bar\pp,\o'}
\ra+A_3(\bar\pp)\la \hat\psi^-_{\kk,\o}\hat\psi^+_{\kk-\hat\pp,\o}
\hat\psi^-_{\kk'-\bar\pp, \o'}
\hat\psi^+_{\kk'+\hat\pp-\bar\pp,\o'} \ra \nn\\
&&+A_4(\bar\pp)\la \hat\psi^-_{\kk,\o}\hat\psi^+_{\kk-\hat\pp,\o}
\hat\psi^-_{\kk', \o'} \hat\psi^+_{\kk'+\hat\pp,\o'}
\ra+H^{4,1}(\kk,\kk',\bar\pp)]\label{2.11aacf} \eea
where $\chi_\e(\bar\pp)$ is a compact support function vanishing
for $\bar\pp=0$, and such that it becomes the identity in the
limit $\e\to 0$;  moreover the functions $A_i(\pp)$ are bounded
and even in $\pp$, and $H^{4,1}$ is a linear combination of the
$R^{4,1}$ functions in \pref{cd1212} with bounded coefficients.

We have now to bound all the sums in the r.h.s. of
\pref{2.11aacf}. Note first that, by parity
\be \int d\bar p_0 d\bar p_+ \chi_\e(\bar\pp) A_4(\bar\pp){\hat
v(\bar p_0,\bar p_+)\over -i \bar p_0+\o''v_F \bar p_+} \la
\hat\psi^-_{\kk,\o}\hat\psi^+_{\kk-\hat\pp,\o} \hat\psi^-_{\kk',
\o'} \hat\psi^+_{\kk'+\hat\pp,\o'}\ra =0\label{sdgfvt5}\ee
Moreover the first term in the r.h.s. of \pref{2.11aacf} verifies
\be |\int d\bar p_0 d\bar p_+ {\hat v(\bar p_0,\bar p_+)\over -i
\bar p_0+v_F\o'' \bar p_+} A_1(\bar\pp)\chi_\e(\bar\pp) \la
\hat\psi^-_{\kk-\bar\pp,\o}\hat\psi^+_{\kk-\hat\pp,\o}
\hat\psi^-_{\kk', \o'} \hat\psi^+_{\kk'+\hat\pp-\bar\pp,\o'} \ra |
\le C\e_{h} {\g^{-3 h}\over Z_h^2}\label{2.24cvcv} \ee
and a similar bound is true for the second and third term.
Finally, as in [16]
\be |\int d\bar p_0 d\bar p_+ {\hat v(\bar p_0,\bar p_+) \over -i
\bar p_0+v_F\o'' \bar p_+} H^{4,1}(\kk,\kk',\bar\pp)|\le C\e_h
{\g^{-3 h}\over Z_h^2} \label{adf} \ee
By inserting
\pref{44},\pref{45},\pref{2.11aacf},\pref{sdgfvt5},\pref{2.24cvcv}
in \pref{sd} we get $\l_h=\l_0+O(U^2)$, which means that the
effective interaction remain close to initial value for any RG
iteration; a contradiction argument shows that this can be true
only if the beta function is asymptotically vanishing: as this
beta function is the same of the 2D model \pref{h} with effective
couplings (30), up to $O(\g^h)$ terms, then \pref{cd11} follows.

\section{Conclusions}

We have shown that the RG flow for a system of spinless fermions
with flat Fermi surface has, in addition to the trivial fixed
point, a line of Luttinger liquid fixed points, corresponding to
vanishing wave function renormalization and anomalous exponents in
the 2-point function; such fixed point is in the invariant set
(30). This makes quantitative the analysis in [10], in which the
existence of a Luttinger fixed point in 2D was postulated on the
basis of bosonization. With respect to previous perturbative RG
analysis,the key novelty is the implementation of WI at each RG
iteration, in analogy to what is done in 1D.

Of course the other effective interactions should cause flows away
from this fixed point. Indeed the situation is somewhat similar to
the 1D (spinning) Hubbard model, in which there is a Luttinger
liquid fixed point in the invariant set
obtained setting all
but the backscattering and umklapp scattering terms equal to zero
(that is the set $g_{1,h}=g_{3,h}=0$ in the g-ology notation, see
[19]), and a flow to strong coupling regime driven by the
backscattering interaction. We can in any case expect, as in 1D,
that even if the Luttinger fixed point in 2D is not stable its
presence has an important role in the physical properties of the
system.

Fermi surfaces with flat or almost flat pieces and no van Hove
singularities are found in the Hubbard model with next to nearest
neighbor interactions or in the Hubbard model close to half
filling, and it is likely that our results can be extended, at
least partially, to such models.  Note however that in such models
the sides of the Fermi surface are not perfectly flat, so that one
expects a renormalization of the shape of the Fermi surface, as in
[3], which is absent in the case of flat sides by symmetry.
Another simplifying property of the model considered here is that
the modulation of the Fermi velocity is taken constant along the
Fermi surface, contrary to what happens in more realistic models;
a momentum dependent Fermi velocity produces extra terms in the
WI, as it is evident from (43) ($v_F$ should be replaced by
$v_F(k'_-)$ in the first line and $v_F(k'_-+p_-)$ in the second
line), and their effect deserves further analysis.


\vskip1cm {\it Acknowledgments} I
am grateful to A.Ferraz and E.Langmnann for very interesting
discussions on the 2D Hubbard model, and to the Schroedinger
institute in Wien where this paper was partly written.

\end{document}